\documentclass[letterpaper, preprint, paper,11pt]{AAS}	% for preprint proceedings

\usepackage{bm}
\usepackage{amsmath}
\usepackage{subfigure}
\usepackage[colorlinks=true, pdfstartview=FitV, linkcolor=black, citecolor= black, urlcolor= black]{hyperref}
\usepackage{overcite}
\usepackage{footnpag}			      	% make footnote symbols restart on each page

\usepackage{amsfonts}
 \newtheorem{assumption}{Assumption}
  \newtheorem{lemma}{Lemma}
 
 \newtheorem{remark}{Remark}
 \newtheorem{problem}{Problem}

 \usepackage{algorithm}
\usepackage[noend]{algpseudocode}

\usepackage{enumitem}
\PaperNumber{23-221}

\begin{document}

\title{Autonomous Satellite Docking via Adaptive Optimal Output Regulation: A Reinforcement Learning Approach}

\author{Omar Qasem\thanks{Ph.D. Candidate, Mechanical Engineering, Florida Institute of Technology, Melbourne, FL, 32901, USA.},  
Madhur Tiwari\thanks{Assistant Professor, Aerospace Engineering, Florida Institute of Technology, Melbourne, FL, 32901, USA.},
\ and Hector Gurierrez\thanks{Professor, Mechanical Engineering, Florida Institute of Technology, Melbourne, FL, 32901, USA.}
}

\maketitle{}

\begin{abstract}
This paper describes an online off-policy data-driven reinforcement learning based-algorithm
to regulate and control the relative position of a deputy satellite in an autonomous satellite
docking problem. The optimal control policy is learned under the framework of output regulation problem and adaptive dynamic
programming (ADP) by considering the continuous-time linearized model of the satellite. The linearized model of relative motion
is used to describe the motion between satellites, and the satellite docking problem is formulated as a linear optimal output regulation problem, in which the feedback-forward optimal controller is used to track a class of references and rejecting a class of disturbances while maintaining the overall system's closed-loop stability. The optimal control problem is presented using a
data-driven reinforcement learning based method to regulate the relative position and velocity
of the deputy to safely dock with the chief. Using the adaptive optimal output regulation framework,
the learned optimal feedback-feedforward gains guarantee optimal transient and steady state
performances without any prior knowledge of the dynamics of the studied system. {The states/input information of the underlying dynamical system are instead used to compute the approximated optimal feedback-feedforward control gain matrices.} Reference tracking and disturbance rejection are achieved in an optimal sense
without using any modelling information of the physics of the satellites. {Simulation results are presented and demonstrate the efficacy of the proposed method.}
\end{abstract}
\section{Introduction}

One of the most important functions of an autonomous system is to be able to follow a specified
path, pertinent to the mission. Trajetory tracking is often accomplished using a combination of a
waypoint path planner and a control system that is used to follow the computed way-points. Path
planning is an extensively researched topic but it is not pertinent to planners can be converted into
3-D trajectories. The control effort required to achieve these maneuvers can then be used by the
controller to follow the trajectory. In the autonomous satellite docking problem, the main goal is achieve the position tracking of the deputy satellite to the chief satellite. The deputy satellite objective is to complete the docking procedure
with the chief. In other words, the relative position/velocity of the deputy needs to follow
some reference input generated by chief. Under this framework, we consider the output regulation
problem to achieve the autonomous satellite docking procedure.

The output regulation problem has gained the consideration and attraction of a wide audience in control systems society since it is a general mathematical formulation to tremendous control problems applications in engineering, biology, satellite clustering and other disciplines; see, for instance, \cite{bonivento2001output, huang2004nonlinear, isidori2003robust, sontag2003adaptation, trentelman2002control} and many references therein. The linear output regulation problem is mainly concerned in designing a control policy to achieve asymptotic tracking of a class of reference inputs, in addition to rejecting nonvanishing disturbances, in which both the reference signals and the disturbances are generated by a class of an autonomous systems, named exosystems. Essentially, the output regulation problem is solved using either feedback-feedforward method, or the internal model principle. In this work, we focus on solving the output regulation problem in the feedback-feedforward scheme.

Solving the output regulation problem by itself cannot guarantee an optimal behaviour for the studied dynamical system, whether in its transient or steady state responses. Dynamic programming (DP) is a backbone in solving optimal control problems. DP was first introduce by Bellman in the early 1950s \cite{bellman1954theory,Bellman1956,Bellman1957}. Bellman principle of optimality is the fundamental idea behind DP, which states that an optimal policy has the property that the following actions must achieve an optimal policy with regard to the state resulting from those previous actions, no matter what previous actions have been \cite{Frank2012optimal}. Based on the theoretical foundation of DP, reinforcement learning \cite{sutton2018reinforcement} and adaptive dynamic programming (ADP) \cite{Lewis2009}, methods have been developed to provide learning-based solutions to optimal control and decision making problems without using the modeling information. 
In particular, ADP has gained numerous attentions recently and been applied to control both continuous-time systems \cite{Lewis2009, jiang2012computational, Gao2014WCICA,Bian2014Auto,bertsekas2015value,Bian2016,Bian2016TAC,zhong2016event,kiumarsi2017optimal,wang2017adaptive,article,doi:10.1137/18M1214147,fong2018dual,rizvi2019reinforcement,9034079,Vamvoudakis2015Auto} and discrete-time systems \cite{wei2015value,6917009,6815973,liu2019h,8685683,Jiang2020FnT}.

\begin{figure}
    \centering
    \includegraphics{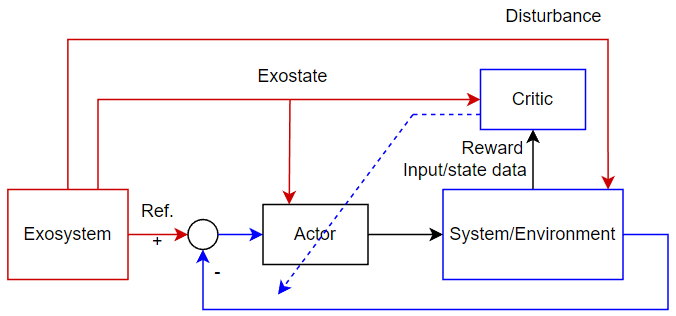}
    \caption{The framework of the learning-based adaptive optimal output regulation}
    \label{fig: actor critic diagram}
\end{figure}

Therefore, different studies have considered combining the theories of adaptive optimal control with the output regulation in order to achieve the adaptive optimal output regulation problem; see \cite{Qasem2022,GAO2022110366,Deng2020,Zhao2021,Gao2017ACC,Gao2015MICNON,Gao2015ACC,Gao2016TAC,Gao2016CDC,Gao2016ACC,Gao2016JINT,Gao2014WCICA,Gao2016Auto} and references therein. Using reinforcement learning and Bellman's principle of optimality \cite{sutton2018reinforcement}, ADP methods
\cite{gao2021reinforcement,he2019adaptive,JiangBook2017,kamalapurkar2018reinforcement,vamvoudakis2012multi,wei2020continuous,yang2021model,Bian2016,bian2021reinforcement,heydari2018stability,
rizvi2019reinforcement,
zhao2020event,QasemHI,GAO2022110366,gao2022learning}, which are essentially based on reinforcement learning,
 are developed such that the agent can learn towards the optimal control policy by interacting with its unknown environment. With this learning framework, see Fig. \ref{fig: actor critic diagram}, and by taking into account the output regulation problem, one can develop an adaptive optimal feedback-feedforward controller which behaves optimally on a long term without the knowledge of the system matrices.

In addition to the asymptotic tracking and disturbance rejection, two minimization problems of a predefined costs function are also considered, in which by solving these minimization problems, the optimal output regulation is achieved. Besides the issue of maintaining the asymptotic tracking, obtaining the full knowledge of the dynamics of the satellite is usually a difficult task, or even impossible. Moreover, the modelling information may not be exact enough which can cause modelling mismatch and therefore, the designed controller may not achieve satisfactory results. To fill in the gap between the output regulation problem and optimality, and overcome the barrier of modelling the physics of the system, a data-driven optimal controller is designed to approximate the feedback and feedforward control gains without the knowledge of the dynamics of the satellite (deputy) using the sate/input information collected along the trajectories of the deputy. 

The contribution of this paper is summarized by the following:
%\begin{enumerate}[label=(\roman*)]
%\item 
(i) We consider the autonomous satellite docking problem based on the ADP and under the framework of the output regulation problem.
%\item 
(ii) The Clohessy-Wiltshire equations are considered wherein the optimal feedback-feedforward control gain matrices are obtained using value iteration.
%\item 
(iii) It is shown the tracking of the deputy to the chief is perfectly achieved in an optimal sense by considering reinforcement learning with the output regulation problem.
%\item 
(iv) To best of our knowledge, this work is the first of its kind to consider ADP strategies and the output regulation concept in autonomous satellite docking applications to regulate the relative position of the deputy to chief.
%\end{enumerate}
\subsubsection*{\textbf{Notations.}}
The operator $|\cdot|$ represents the Euclidean norm for vectors and the induced norm for matrices. $\mathbb{Z}_{+}$ denotes the set of nonnegative integers. The Kronecker product is represented by $\otimes$, and the block diagonal matrix operator is denoted by $\textrm{bdiag}$. 
$I_n$ denotes the identity matrix of dimension $n$ and $0_{n\times m}$ denotes a $n\times m$ zero matrix. 
$\text{vec}(A) = [a_1^\textrm{T},a_2^\textrm{T},...,a_m^\textrm{T}]^\textrm{T}$, where $a_i \in \mathbb{R}^n$ is the $i^{\text{th}}$ column of $A \in \mathbb{R}^{n\times m}$. 
For a symmetric matrix $P=P^\textrm{T} \in \mathbb{R}^{m\times m},$ $\text{vecs}(P)=\left[p_{11},2p_{12},...,2p_{1m},p_{22},2p_{23},...,2p_{m-1,m},p_{mm}\right]^\textrm{T}\in \mathbb{R}^{\frac{1}{2}m(m+1)}$. 
$P\succ(\succeq)0$ and $P\prec(\preceq)0$ denote the matrix $P$ is positive definite (semidefinite) and negative definite (semidefinite), respectively. For a column vector $v\in \mathbb{R}^n$, ${\rm vecv}(v)$$=$$[v_1^2,v_1 v_2,\cdots,v_1v_n,v_2^2,v_2v_3,\cdots,v_{n-1}v_n,v_n^2]^T\in\mathbb{R}^{\frac{1}{2}n(n+1)}$. For a matrix $A\in \mathbb{R}^{n\times n}$, $\sigma(A)$ denotes the spectrum of $A$. For any $\lambda \in \sigma(A)$, $\text{Re}(\lambda)$ represents the real part of the eigenvalue $\lambda$.

\section{Objective and Methodology}
The objective of this paper is to study and analyze the docking scenario of a deputy satellite into a chief satellite such that the process is done with guaranteed stability, and maintaining the asymptotic tracking of the deputy to the chief. Moreover, the method should not rely on the prior knowledge of the physics of the system.Therefore, the learning method is carried by the adaptive optimal output regulation, in which the optimal control policy are learned using ADP.

To begin with, consider the following continuous-time linear system described by 
\begin{align}\label{eq: exosystem}
\dot{v}&=Ev,\\\label{eq: x-system}
    \dot{x}&=Ax+Bu+Dv\\\label{eq: e system}
    e&=Cx+Fv,
\end{align}
where the vector $x\in\mathbb{R}^{n}$ is the state, $u\in\mathbb{R}^{m}$ is the control input, and $v\in\mathbb{R}^q$ stands for the exostate of an autonomous system \eqref{eq: exosystem}. The vector $e\in\mathbb{R}^{p}$ represents the output tracking error. The matrices $A\in\mathbb{R}^{n\times n}$, $B\in\mathbb{R}^{n\times m}$, $D\in\mathbb{R}^{n\times q}$, $C\in\mathbb{R}^{p\times n}$, and $F\in\mathbb{R}^{p\times q}$ are real matrices with the pair $(A,B)$ are assumed to be unknown.

In our case, the relative motion between a 'Chief' and a 'Deputy' which are in close vicinity of each other is defined by the Clohessy-Wiltshire (CW) equations \cite{schaub2003analytical}. Relative accelerations in Cartesian coordinates are given by:
\begin{align}\label{eq: CW 1}
    \Ddot{\textbf{x}}-2\bar n\dot{\textbf{y}}-3\bar{n}^2 \textbf{x}&=0,\\\label{eq: CW 2}
    \ddot{\textbf{y}}+2\bar{n}\dot{\textbf{x}}&=0,\\\label{eq: CW 3}
    \ddot{\textbf{z}}+\bar{n}^2\textbf{z}&=0,
\end{align}
where $(\textbf{x},\textbf{y},\textbf{z})$ represent the relative position of the two satellites in the orthogonal Cartesian coordinate system and $\bar n$ is the mean orbital rate. The vector components are taken in the rotating chief Hill frame. The advantage of using Hill frame coordinates is that the physical relative orbit dimensions are immediately apparent from these coordinates. The $(\textbf{x},\textbf{y})$ coordinates define the relative orbit motion in the chief orbit plane. The $\textbf{z}$ coordinate defines any motion out of the chief orbit plane. The following assumptions are taken into account such \eqref{eq: CW 1}-\eqref{eq: CW 3} can be held.
\begin{assumption} The relative distance between the chief and the deputy is much smaller than the orbit radius $r$.
\end{assumption}
\begin{assumption}
The relative orbit is assumed to be circular.
\end{assumption}

To begin with, we define the state space vector $x$ as $x=\left[\textbf{x},~\textbf{y},~ \textbf{z},~\dot{\textbf{x}},~\dot{\textbf{y}},~ \dot{\textbf{z}}\right]^\textrm{T}$, and the control input vector as $u=\left[T_{1},~T_{2},~T_{3}\right]^\textrm{T}$, such that the thrusters are available to control the deputy in any of the three directions. We can then write eqs \eqref{eq: CW 1}-\eqref{eq: CW 3} in the form of \eqref{eq: exosystem}-\eqref{eq: e system}.

Some general assumptions are also considered when solving the output regulation as follows.
\begin{assumption}\label{assumption: controllability}
The pairs ($A$,$B$) and ($C$,$A$) are stabilizable and observable, respectively.
\end{assumption}

\begin{assumption}\label{assumption: rank}
{\rm rank}$\left(\begin{bmatrix}
A-\lambda I_n & B\\C&0
\end{bmatrix}\right) = n +p,\,\forall\lambda \in \sigma (E)$.
\end{assumption}
In order to solve the optimal output regulation problem, {two optimization} problems need to be addressed. {The static optimization Problem \ref{Problem 1} is solved in order to find the optimal solution $(X^\star,U^\star)$ to the regulator equations \eqref{regeq1}-\eqref{regeq2}. While the dynamic optimization problem described in Problem \ref{Problem 2} is solved to find the optimal feedback control policy.} Both problems are stated as follows:

 \begin{problem}\label{Problem 1}
  \begin{align}
      \min_{(X,U)}{\rm Tr}&(X^{\rm T}\bar{Q}X+U^{\rm T}\bar{R}U), \label{min_Tr}\\
   {\rm subject~ to~~    }  XE&=AX+BU+D,\label{regeq1}\\
     0&=CX+F,\label{regeq2} 
  \end{align}
  where $\bar Q=\left(\bar Q\right)^{\rm T}\succ0$ and $\bar R=\left(\bar R\right)^{\rm T}\succ0$.
   \end{problem}
    \nocite{ABDULLAH20195541}
  Based on Assumption \ref{assumption: rank}, the solvability of the regulator equations defined by \eqref{regeq1}-\eqref{regeq2} is guaranteed and the pair $(X,U)$ exist for any matrices $D$ and $F$; see \cite{huang2004nonlinear}. 
  Additionally, the solution to Problem \ref{Problem 1}, i.e., $(X^\star,U^\star)$ is unique, which will guarantee that the feedforward control policy obtained using $(X^\star,U^\star)$ is also unique and optimal.  
  \begin{problem}\label{Problem 2}
  \begin{align}
      \min_{\bar{u}_i}  \int_{0}^{\infty} &\left(\bar{x}^{\rm T}Q\bar{x}+\bar{u}^{\rm T}{R}\bar{u}\right) ~{\rm d}t, \\\label{eq: error system 1}
      {\rm~subject~ to~~ } \dot{\bar{x}}&=A\bar{x}+B\bar{u},\\\label{eq: error system 2}
      e&=C\bar{x},
  \end{align}
  where $Q=\left(Q\right)^{\rm T} \succeq 0, R =\left(R\right)^{\rm T} \succ0,$ with $\left(A,\sqrt{Q}\right)$ being observable. The equations \eqref{eq: error system 1}-\eqref{eq: error system 2} form the error system with $\bar{x}:=x-Xv$ and $\bar{u}:=u-Uv$.
  \end{problem}
  
Note that if the deputy's dynamics in \eqref{eq: x-system} are perfectly known, one can develop the optimal controller in the following form.
\begin{align}\label{eq: ui*}
    u^\star(K^\star,L^\star)=-K^\star x+L^\star v,
\end{align} where $K^\star=R^{-1}B^\textrm{T}P^\star$, and $P^\star $ is the unique solution of the following albegraic Ricatti equation (ARE)
\begin{align}
    A^\textrm{T}P^\star+P^\star A+Q-P^\star BR^{-1}B^\textrm{T}P^\star=0.\label{eq: ARE}
\end{align}

The solutions to the regulator equations \eqref{regeq1}-\eqref{regeq2}, i.e., $(X,U)$, form the optimal feedforwad gain matrix such that \begin{align}\label{eq: Li*}
    L^\star&=U+K^\star X.
\end{align}

It is remarkable that equation \eqref{eq: ARE} is nonlinear in $P^\star $. Therefore, different iterative methods have considered to solve the ARE iteratively, including policy iteration (PI) and value iteration (VI). The following lemma shows the convergence of \eqref{eq: ARE} in the sense of the PI method.
\begin{lemma}[\hspace{-0.2pt}\cite{Kleiman1968}]
Let $K_{0} \in \mathbb{R}^{m \times n}$ be a stabilizing feedback gain matrix, the matrix $P_{k}=(P_{k})^{\rm T}\succ0$ be the solution of the following equation  
\begin{align}\label{eq: Policy evaluation}
   P_{k}(A-BK_{k-1})+(A-&BK_{k-1})^{\rm T} P_{k} +Q+K_{k-1}^{\rm T} R K_{k-1}=0,
\end{align} 
 and the control gain matrix $K_{k}$, with $k=1,2,\cdots,$ are defined recursively by
\begin{align}\label{eq: Policy improvement}
    K_{k}=R^{-1} B^\textrm{T} P_{k-1}.
\end{align} 
Then the following properties hold for any $k\in \mathbb{Z}_{+}$. 
\begin{enumerate}%[label=\roman*.]
    \item The matrix $A-BK_{k}$ is Hurwitz.
    \item $P^\star \preceq P_{k} \preceq P_{k-1}$.
    \item $\underset{{k\rightarrow \infty}}\lim K_{k} = K^\star ,\; \underset{{k\rightarrow \infty}}\lim P_{k}=P^\star $.
\end{enumerate}
\end{lemma}  

It is notable that an initial stabilizing control policy is required to initiate the learning process of PI. In this paper we consider an iterative reinforcement learning method based on VI to solve $P^\star $, wherein the solvability of the VI is considered under ADP scheme. We consider the use of VI since no initial stabilizing control policy is required to initiate the learning process. This gives VI an advantage over PI since obtaining the prior knowledge of an initial stabilizing control policy is a stringent requirement and may be impossible to obtain, especially when the system dynamics are not available or are not known perfectly. The iterative process of VI to find the optimal control policy is done by repeating the value update step until the value function converges to its optimal value. In the following sections, we show in further details the use of VI to solve our problem.

\section{Model-Based Value Iteration% for Output Regulation Problem
}
Throughout this section, the value iteration is used such that the value matrix is iteratively updated until the value matrix converges within a predefined condition.
{To begin with}, $\{B_r\}_{r=0}^{\infty}$ is defined as a collection of nonempty interiors bounded sets, which satisfies 
\begin{align*}
B_r \subset B_{r+1} \in \mathcal{J}_+^n , \; r\in \mathbb{Z}_{+},\; \lim_{r \rightarrow \infty} B_r  = \mathcal{J}^{n}_{+},\end{align*} and $\varepsilon>0$ is a small constant selected as a threshold. In addition, select a deterministic sequence $\{\epsilon_{k}\}_{k=0}^{\infty}$ such that the following conditions are satisfied: \begin{align}\label{eq: ek condition}\epsilon_k>0,\;\;\sum_{k=0}^{\infty}\epsilon_k=\infty,\;\;\lim_{k \rightarrow 0}\epsilon_k = 0.\end{align} 

As mentioned earlier, the VI is different from the policy iteration, described by \eqref{eq: Policy evaluation}-\eqref{eq: Policy improvement} in the sense that an initial stabilizing control policy is not required. Instead, the learning process is initiated with an arbitrary value matrix $P_{0}=(P_{0})^\textrm{T}\succ 0$. In the following, the model-based VI algorithm is given, in which the system matrices $(A,B)$ are used to learn the optimal control policy, based on the results in \cite{Bian2016}.
\begin{algorithm}%[H]
\begin{algorithmic}[1]
\caption{Model-based Value Iteration}
% \Repeat
\State Select a small constant
$\varepsilon>0$, and
{$P_{0}=(P_{0})^\textrm{T}\succ0$}.
\State $k,r \gets 0$.
\Repeat
 \State ${\Tilde{P}_{k+1}\gets P_{k}+\epsilon_k (P_{k}A+A^{\textrm{T}}P_{k}+ {  %C^\textrm{T}
 Q%C
 }} {-P_{k} B R^{-1} B^{\textrm{T}} P_{k})}$
\If{$\Tilde{P}_{k+1}\notin B_{r}$}
{$P_{k+1}\leftarrow P_{0},~ r\leftarrow r+1$}.
\Else 
{ $P_{k+1}\gets\Tilde{P}_{k+1}$} 
\EndIf {\textbf{endif}}
\State $k\gets k+1$
\Until $|\Tilde{P}_{k}-P_{k-1}|/\epsilon_{k-1}\prec\varepsilon$
\State $k^\star \gets k$
\State Find the pair $(X,U)$ from \eqref{regeq1}-\eqref{regeq2}.
\State $L_{k\star}\gets U + K_{k^\star}X$
\State Obtain the optimal controller {using} 
 $   u^\star=-K_{k^\star}x+L_{k^\star}v.$
\label{model-based VI Algorithm}
\end{algorithmic}
\end{algorithm}
% {\begin{theorem}
% \myr The sequences $\{P_{k}\}_{k=0}^\infty$ and $\{K_{k}\}_{k=1}^\infty$ computed by Algorithm \ref{model-based VI Algorithm} satisfy the inequalities $|P_{k^*}-P^*|\leq \varepsilon$ and $|K_{k^*}-K^*|\leq \varepsilon$, where $\varepsilon>0$ is a small threshold and $k^*\in\mathbb{Z}_+$.
% \end{theorem}}
\begin{remark}
    It is noteworthy to mention that if the bound of $P^\star$ is known in prior, i.e., $|P^\star|<\gamma$, then one can fix $B_r$ to $B_r=\gamma$. 
\end{remark}

\section{Data-Driven Value Iteration for Output Regulation Problem}
From the previous section, it is notable that the model-based VI requires the full knowledge of the system matrices $(A,B)$. In practice, obtaining these matrices may not be easy when considering higher order and more complex systems. In this section, we consider a data-driven VI method in which the optimal control policy is obtained without relying on the dynamics or the physics of the system, but the data (state/input information) collected along the trajectories of the underlying dynamical system are used to learn an approximated optimal control policy. 

Considering the $x-$system in \eqref{eq: x-system}, define $\bar{x}_{j}=x - X_{j}v$ for $0\leq j\leq h +1$, where $X_{0}=0_{n \times q}$, $X_{j}\in \mathbb{R}^{n \times q}$ so that $C X_{1}+F=0$. The matrices $X_{j}%\in \mathbb{R}^{n_i \times q}
$ for $2\leq j\leq h +1$, where $h = (n - p)q$ is the null space dimension of $I_{q}\otimes C$, are selected such that the basis for $\text{ker}(I_{q}\otimes C)$ are formed by all the vectors $\text{vec}(X_{j})$. With the above definitions along with \eqref{eq: exosystem}--\eqref{eq: x-system}, the following differential equation is then obtained.
\begin{align} \label{xbar equation for VI}
    \dot{\bar x}_{j}&= A x + B u +(D-X_{j}E)v\\ 
    \label{xbar equation for PI}
    %{ \dot{\bar x}_{j}}
    &{ = A_{k} \bar{x}_{j} +B (K_{k} \bar x_{j}+u)+(D - S (X_{j}))v}
\end{align}
where the Sylvester map $S:\mathbb{R}^{n \times q}\rightarrow \mathbb{R}^{n\times q}$ satisfies $S(X)=XE-AX$, $\forall$ $X\in \mathbb{R}^{n\times q}$, and $A_{k}=A - B K_{k}$.
For any two vectors {$a(t)\in\mathbb{R}^{\boldsymbol{n}},b(t)\in\mathbb{R}^{\boldsymbol{m}}$}, and a sufficiently large $\rho\in\mathbb{Z}_+$, the following matrices are defined.
    \begin{align}
    \delta_b&=\left[\text{vecv}({ b})|_{t_0}^{t_1},\text{vecv}({ b})|_{t_1}^{t_2} \cdots,\text{vecv}({ b})|_{t_{\rho -1}}^{t_\rho}\right]^\textrm{T}{\in\mathbb{R}^{\rho\times {\boldsymbol{m}(\boldsymbol{m}+1)}/{2}}},\nonumber\\
%   \delta_b&=\Bigl[\text{vecv}(b(t_1))-\text{vecv}(b(t_0)), \cdots,\nonumber\\&\hspace{88pt}\text{vecv}(b(t_\rho))-\text{vecv}(b(t_{\rho-1}))\Bigr]^\textrm{T}\nonumber\\&~~~~~~~~~~~~~~~~~~~~~~~~~~~~~~~~~~~~~~~~~~~~~~~{\new\in\mathbb{R}^{\rho\times {\boldsymbol{m}(\boldsymbol{m}+1)}/{2}}},\nonumber\\
    \Gamma_{a,b}&=\begin{bmatrix}\int_{t_0}^{t_1}a\otimes b\;\textrm{d}\tau,\int_{t_1}^{t_2}a\otimes b\;\textrm{d}\tau,\cdots,\int_{t_{\rho-1}}^{t_\rho}a\otimes b\;\textrm{d}\tau\end{bmatrix}^\textrm{T}
  {\in\mathbb{R}^{\rho\times \boldsymbol{n}\boldsymbol{m}}}.\nonumber 
     \\
      {\mathbb{I}_{a}}&{ =\begin{bmatrix}\int_{t_0}^{t_1}\text{vecv}(a)\textrm{d}\tau,\int_{t_1}^{t_2}\text{vecv}(a)\textrm{d}\tau \cdots,\int_{t_{\rho -1}}^{t_\rho}\text{vecv}(a)\textrm{d}\tau\end{bmatrix}^\textrm{T}\nonumber}{{\in\mathbb{R}^{\rho\times {\boldsymbol{m}(\boldsymbol{m}+1)}/{2}}}.\nonumber}
    \end{align}

Consider the Lyapunov candidate $V_{k}(\bar x_{j})=\bar x_{j}^\textrm{T} P_{k} \bar x_{j}$, where $k\in \mathbb{Z}_{+}$. By taking the time derivative of $V_{k}(\bar x_{j})$ along with %\eqref{xbar equation for VI} 
\eqref{xbar equation for PI}, with some mathematical manipulations and rearrangements, one obtains the following.%\eqref{VI Lyapunov}.
\begin{align}%\label{Vdot}
    \dot V_k(\bar x_{j})&=\dot{ \bar x}_{j}^\textrm{T}P_{k}\bar x_{j}+\bar x_{j}^\textrm{T}P_{k}\dot{ \bar x}_{j}\nonumber\\
    \label{VI Lyapunov}
    &=\bar x_{j}^\textrm{T}(H_{k})\bar x_{j}+2u^\textrm{T}RK_{k+1}\bar x_{j}+2v^\textrm{T}(D - S(X_{j}))^\textrm{T}P_{k}\bar x_{j}
\end{align}
where $H_{k}=A^\textrm{T}P_{k}+P_{k}A$.\\

By taking the integral of \eqref{VI Lyapunov} over $[t_0, t_s]$, {where $\left\{t_l\right\}_{l=0}^{s}$ with $t_l=t_{l-1}+\Delta t,$ $\Delta t>0$ is a strictly increasing sequence}, the result can be written in the following Kronecker product representation.
\begin{align}\label{VI data-driven}
    &{\Theta_{j}}\begin{bmatrix}
    {\text{vecs}}(H_{k})\\\text{vec}(K_{k+1})\\\text{vec}((D - S(X_{j}))^{\textrm{T}}P_{k})
    \end{bmatrix}=\delta_{\bar x_{j} , \bar x_{j}}\text{vecs} (P_{k})\hspace{-0.2em}
\end{align}
where ${\Theta_{j}} = \begin{bmatrix}
    %\Gamma_{\bar x_{j},\bar x_{j}}
{\mathbb{I}_{\bar{x}_{j}}}
    ,& 2\Gamma_{\bar x_{j},{u}}(I_{n} \otimes R),&2\Gamma_{\bar x_{j},{v}}
    \end{bmatrix} $. If $\Theta_j$ is full column rank, the solution of \eqref{VI data-driven} is obtained in the sense of least square error by using the pseudo-inverse of $\Theta_j$, i.e., $\Theta_j^\dagger=\left(\Theta_j^\textrm{T}\Theta_j\right)^{-1}\Theta_j^\textrm{T}$. The full column rank condition of $\Theta_j$ is satisfied by the following lemma.  
    \begin{lemma}\label{rank lemma}
For all $j\in \mathbb{Z}_{+}$, if there exist a $s'\in \mathbb{Z}_{+}$ such that for all $s>s'$ the following rank condition is satisfied
\begin{align}\label{eq: rank}
{\rm rank}\left(\begin{bmatrix}%\Gamma_{\bar x_{ij},\bar x_{ij}}
\mathbb{I}_{\bar x_{j}}, \Gamma_{\bar x_{j} ,u}, \Gamma_{\bar x_{j},v}
\end{bmatrix}\right)=\frac{n (n+1)}{2}+(m+q)n
\end{align} for any increasing sequence $\{t_l\}_{l=0}^{s}$, $t_l=t_{l-1}+\Delta t,$ $\Delta t>0$, then the matrix {$\Theta_{j}$} has full column rank, $\forall\;k\in \mathbb{Z}_{+}.$
\end{lemma}

Lemma \ref{rank lemma} shows that if \eqref{eq: rank} is satisfied, the existence and uniqueness of the solution of \eqref{VI data-driven} is guaranteed, {where the solution can be obtained using the pseudo-inverse of {$\Theta_{j}$}.}

{\begin{remark}
The matrix $\Theta_{j}$ is fixed for all $k\in\mathbb{Z}_+$ and does not require to be updated at each iteration $k$.
\end{remark}}

To this end the value matrix is updated using stochastic approximation by
\begin{align*}
	{P}_{k+1}\gets P_{k} + \epsilon_k(H_{k} + Q -(K_{k+1})^\textrm{T}RK_{k+1})
\end{align*} where $\epsilon_k$ satisfies \eqref{eq: ek condition}, until the condition $|P_{k}-P_{k-1}|/\epsilon_k\leq{\varepsilon}$ is satisfied, where ${\varepsilon}>0$ is a small threshold. By that, it is guaranteed that the obtained control policy is close enough to the actual optimal one.

 The data-driven VI algorithm can now be introduced. It is presented in Algorithm \ref{Algorithm:HI Data-driven}.
\begin{algorithm}
\caption{Data-Driven Value Iteration for Optimal Output Regulation}
\begin{algorithmic}[1]
\State Choose a small threshold constant ${\varepsilon}>0$ and  {$P_{0}=(P_{0})^\textrm{T} \succ 0$}.
%\Repeat
\State Compute {the} matrices $X_{0}, X_{1},\cdots, X_{h+1}$.
\State Choose an arbitrary $K_{0}$, not necessarily stabilizing, and employ $u_{0}=-K_{0}x+\eta$, with $\eta$ being an exploration noise over $[t_0,t_s]$. 
\State 
$j\gets 0.$
\Repeat
\State 
%\hspace{-8.75mm}
Compute $\mathbb{I}_{\bar{x}_{j}}\text{, }\Gamma_{\bar x_{j}  u} {\text{, and }} \Gamma_{\bar x_{j} v}$ {while} satisfying \eqref{eq: rank}. % is satisfied.
\State 
$j\gets j+1.$
\Until $j=h+2$
\State $k\gets 0$, $j\gets 0$, $r\gets 0$.
\Repeat
\State Solve $H_{k}$ and $K_{k+1}$ from \eqref{VI data-driven}.
\State $%\hspace{-20pt}
{\Tilde{P}_{k+1}\gets P_{k} + \epsilon_k(H_{k} + Q -(K_{k+1})^\textrm{T}RK_{k+1})}$
\If{$\Tilde{P}{_{k+1}}\notin B_r$} 
% \State
{$P_{k+1}\gets P_{0}$, $r\gets r+1$.}\Else
%\State 
{ $P_{k+1}\gets \Tilde{P}_{k+1}$}
\EndIf \textbf{end if}
\State $k\gets k+1$
\Until $|P_{k}-P_{k-1}|/\epsilon_{k-1}<{\varepsilon}$ 
\State $k\gets k^*, j\gets 1$.
\Repeat
\State From \eqref{VI data-driven}, solve $S({X}_{j})$.
%\State 
$j\gets j+1.$
\Until $j=h +2$

From Problem \ref{Problem 1}, {find} $(X^\star,U^\star)$ using online data.
\State $L_{k\star}\gets U^\star + K_{k^\star}X^\star$
\State Obtain the suboptimal controller {using} 
%(\ref{eqn-observer}) and %(\ref{eq: suboptimal controller}).
\begin{align}\label{eq: suboptimal controller}
    u^\star&=-K_{k^\star}x+L_{k^\star}v.
\end{align}
%\State $i\gets i+1$
%\Until $i=N+1$
\end{algorithmic}
\label{Algorithm:HI Data-driven}
\end{algorithm}

If \eqref{eq: rank} is satisfied, it is guaranteed that the sequences $\{{P_{k}}\}_{k=0}^{\infty}$ and $\{{K_{k}}\}_{k=1}^{\infty}$ learned by Algorithm \ref{Algorithm:HI Data-driven} converge respectively to $P^\star$ and $K^\star$. It is worth mentioning that the proposed VI Algorithm \ref{Algorithm:HI Data-driven} is an off-policy learning algorithm. Since the value function in VI is increasing, the increasing sequence of $\{P_k\}_{k=0}^\infty$ will not affect the trajectories of the system during the learning period.

{\begin{remark}
An exploration noise is added to the input of the system \eqref{eq: x-system}--\eqref{eq: e system} during the learning process of Algorithm \ref{Algorithm:HI Data-driven}. Such an input is chosen to satisfy the rank condition \eqref{eq: rank}---which is similar to the condition of persistent excitation. The noise selected can be a random noise or a summation of sinusoidal signals with distinct frequencies, see \cite{sutton2018reinforcement,Bian2016,jiang2012computational} and references therein.
\end{remark}}

\section{Implementation}
In this section, the simulation of the autonomous satellite docking is implemented using MATLAB, and the results are presented. %will be used to demonstrate the effectiveness of solving the autonomous satellite docking under the framework of output regulation problem. 
The system to be considered is assumed to be under J2 Oblateness perturbation, which is modeled as a disturbance injected into the system. The dynamics considered in this paper are based on the work done in \cite{schweighart2001development}. In this we consider the disturbances to be created by the by the exosystem. The system is described in the following form:
\begin{align}\label{eq: CW1 J2}
    \Ddot{\textbf{x}}-2\bar{n}c\dot{\textbf{y}}-(5c^2-2)\bar{n}^2\textbf{x}&=-3\bar{n}^2J_{2}\frac{R_{e}^2}{r_{\rm ref}}(\frac{1}{2}-\frac{3\sin^2{(i)} \sin^2{(\bar{n}ct)} }{2}-\frac{1+3\cos{(2i)}}{8}),\\
     \ddot{\textbf{y}}+2\bar{n}\dot{\textbf{x}}&=-3\bar{n}^2J_{2}\frac{R_{e}^2}{r_{\rm ref}}\sin^2{(i)}\sin^2{(\bar{n}ct)}\cos{(\bar{n}ct)},\\\label{eq: CW3 J2}
    \ddot{\textbf{z}}+\bar{n}^2\textbf{z}&=-3\bar{n}^2J_{2}\frac{R_{e}^2}{r_{\rm ref}}\sin{(i)}\sin{(\bar{n}ct)}\cos{(i)},
\end{align}where $\bar{n}$ is the mean orbital rate, $R_e$ is the radius of the earth, $r_{\rm ref}$ is the position of the reference orbit, $t$ is the time, and $i$ is the angle of incidence.
The system in \eqref{eq: CW1 J2}-\eqref{eq: CW3 J2} can be reformulated and described in the following form: \begin{align}\label{eq: xddot}
\Ddot{\textbf{x}}-2\bar{n}c\dot{\textbf{y}}-(5c^2-2)\bar{n}^2\textbf{x}&=-3\bar{n}^2J_{2}\frac{R_{e}^2}{r_{\rm ref}}\mathcal{Q}_1,\\\label{eq: yddot}
\ddot{\textbf{y}}+2\bar{n}\dot{\textbf{x}}&=-3\bar{n}^2J_{2}\frac{R_{e}^2}{r_{\rm ref}}\mathcal{Q}_2,\\\label{eq: zddot}
\ddot{\textbf{z}}+\bar{n}^2\textbf{z}&=-3\bar{n}^2J_{2}\frac{R_{e}^2}{r_{\rm ref}}\mathcal{Q}_3,
\end{align}
where $c\equiv\sqrt{1+\textbf{s}}$ with $\textbf{s}=\frac{3J_2R_e^2}{8{r_{\rm ref}}^2}(1+3\cos{2i})$, $\mathcal{Q}_1$, $\mathcal{Q}_2$ and $\mathcal{Q}_3$ are disturbances generated by the exosystem with $|\mathcal{Q}_1|\leq 1$, $|\mathcal{Q}_2|\leq 1$ and $|\mathcal{Q}_3|\leq 1$. Therefore, the above equations can be transformed in the form of \eqref{eq: exosystem}-\eqref{eq: e system} by assuming sinusoidal signals are generated by the exosystem \eqref{eq: exosystem} in addition to the tracking signals. 
{Based on \eqref{eq: xddot}--\eqref{eq: zddot}, the system matrices can be found as follows:
\begin{align}
    A=\begin{bmatrix}
        0&0&0&1&0&0\\
        0&0&0&0&1&0\\
        0&0&0&0&0&1\\
        (5c^2-2)\bar{n}^2&0&0&0&2\bar{n}c&0\\
        0&0&0&-2\bar{n}&0&0\\
        0&0&\Bar{n}^2&0&0&0
 \end{bmatrix},~B=\begin{bmatrix}
     0&0&0\\
     0&0&0\\
     0&0&0\\
     1&0&0\\
     0&1&0\\
     0&0&1
 \end{bmatrix}
\end{align}

The performed simulations are summarized in the following steps:
\begin{enumerate}
    \item An essentially bounded input is applied to deputy satellite along with a non-stabilizing control policy.
    \item State, input and exosystem information are collected along the trajectories of the system described in \eqref{eq: exosystem}-\eqref{eq: x-system} for the time interval $[0,10](s)$.
    \item The optimal control problem is obtained by solving Problem 2, wherein an optimal state feedback gain matrix is obtained.
    \item The output regulation problem is solved by solving Problem 1, wherein the output regulation is then achieved.
    \item The adaptive optimal output regulation is achieved by applying the optimal feedback-feedforward matrix designated in \eqref{eq: Li*} 
\end{enumerate}
The proposed approach shown in Algorithm \ref{Algorithm:HI Data-driven} is used to learn the optimal feedback-feedforward control policy to regulate the relative positions. In addition. Instead of using the modelling information of the system, we use the online collected data to learn the optimal control policy, which removes the stringent requirement of knowing the exact physics of the studied system. The data collection and learning is set to be in the interval $[0,25] (s)$. Last but not least, besides achieving the asymptotic tracking of the exosystem signals, we are also able to achieve rejection for class of disturbances generated by the exosystem, with minimizing a predefined cost function. 
{For simulation purposes, we assume the reference signal and the disturbances are generated by the exosystem with the matrix $E$ defined as follow 
\begin{align}
E&=\textrm{bdiag}\left(\begin{bmatrix}
0&0.1\\-0.1&0\end{bmatrix},\begin{bmatrix}
0&0.2\\-0.2&0\end{bmatrix},\begin{bmatrix}
0&0.3\\-0.3&0\end{bmatrix},
\begin{bmatrix}
0&0.4\\-0.4&0\end{bmatrix}\right)
% \begin{bmatrix}
%      0&1&0&0&0&0&0&0\\
%     -1&0&0&0&0&0&0&0\\
%      0&0&0&2&0&0&0&0\\
%      0&0&-2&0&0&0&0&0\\
%      0&0&0&0&0&3&0&0\\
%      0&0&0&0&-3&0&0&0\\
%      0&0&0&0&0&0&0&4\\
%      0&0&0&0&0&0&-4&0
% \end{bmatrix}
\end{align}
The rest of the matrices are shown below, where $Dv$ represents the disturbances applied to the system, and $-Fv$ is the tracking signal.
\begin{align}
C&=\begin{bmatrix}
    1&0&0&0&0&0\\
    0&1&0&0&0&0\\
    0&0&1&0&0&0
\end{bmatrix},~F=\begin{bmatrix}
    1&0&0&0&0&0&0&0\\
    0&1&0&0&0&0&0&0\\
    0&0&1&0&0&0&0&0
\end{bmatrix},\\
D&=\begin{bmatrix}
0&0&0&0&0&0&0&0\\
0&0&0&0&0&0&0&0\\
0&0&0&0&0&0&0&0\\
0&0&-3\bar{n}^2J_{2}\frac{R_{e}^2}{r_{\rm ref}}&0&0&-3\bar{n}^2J_{2}\frac{R_{e}^2}{r_{\rm ref}}&0&0\\
    0&0&0&0&-3\bar{n}^2J_{2}\frac{R_{e}^2}{r_{\rm ref}}&0&0&0\\
    -3\bar{n}^2J_{2}\frac{R_{e}^2}{r_{\rm ref}}&0&0&0&0&0&-3\bar{n}^2J_{2}\frac{R_{e}^2}{r_{\rm ref}}&0
\end{bmatrix}
\end{align}
}}
The cost function matrices are considered to be $Q=0.05I_6$ and $R=10I_3$. $B_r=10(r+1)$ and $\epsilon_k=\frac{1}{k}$. The CW parameters are chosen same to those used in \cite{schweighart2001development}, where $r_{\textrm{ref}}=7000~km$ and $\Bar{n}=0.00108~1/s$. The altitude of the chief with respect to the center of the earth is chosen to be 6776 km (Low-Earth Orbit). 
\section{Results}
\begin{figure}[H]
    \centering
    \includegraphics[width=0.8\linewidth]{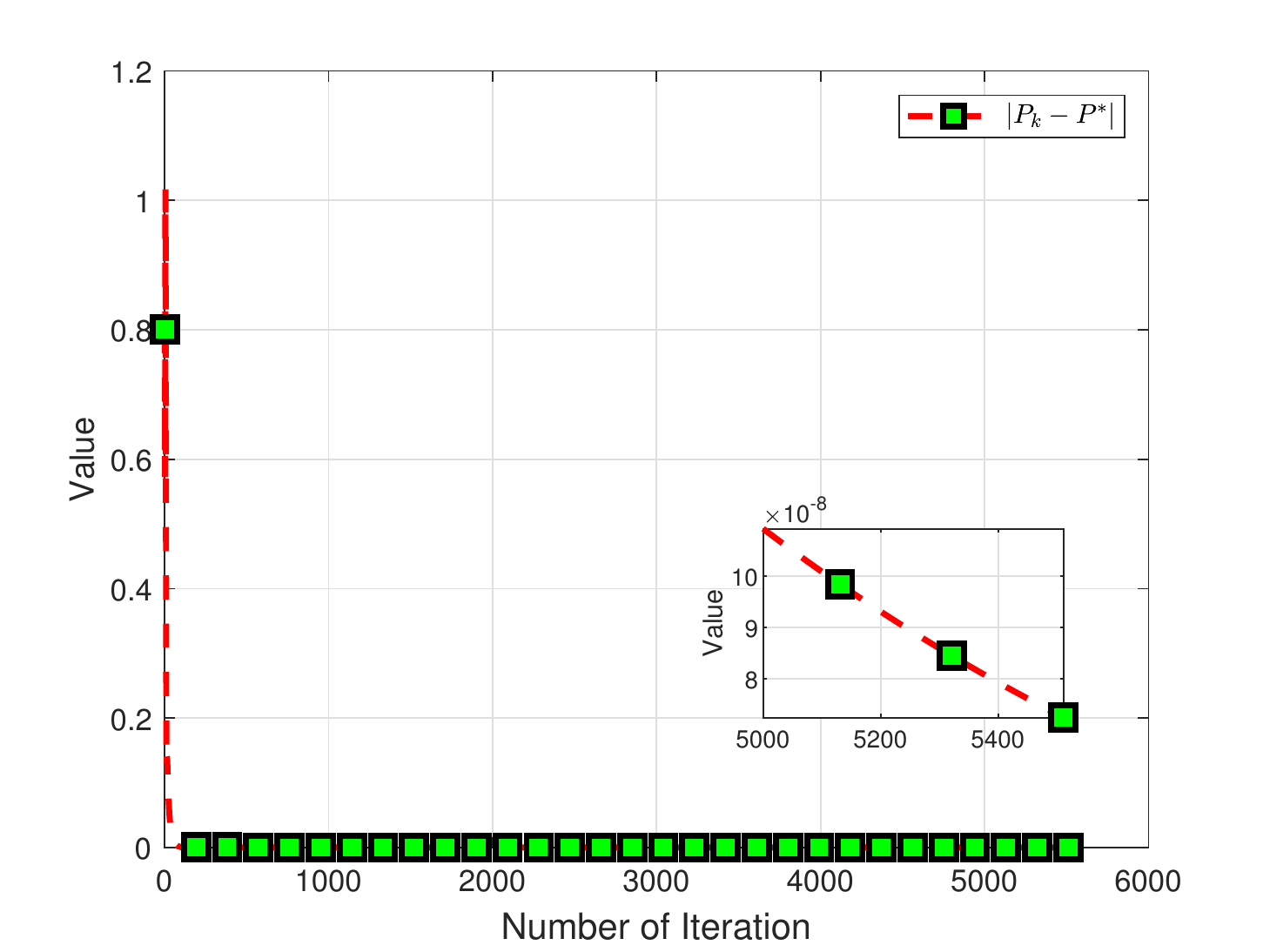}
    \caption{The norm of difference between the learned value matrix $P_k$ and $P^\star$ at each iteration $k$ under value iteration Algorithm \ref{Algorithm:HI Data-driven}}
    \label{fig: errorP}
\end{figure}

\begin{figure}[H]
    \centering
    \includegraphics[width=0.8\linewidth]{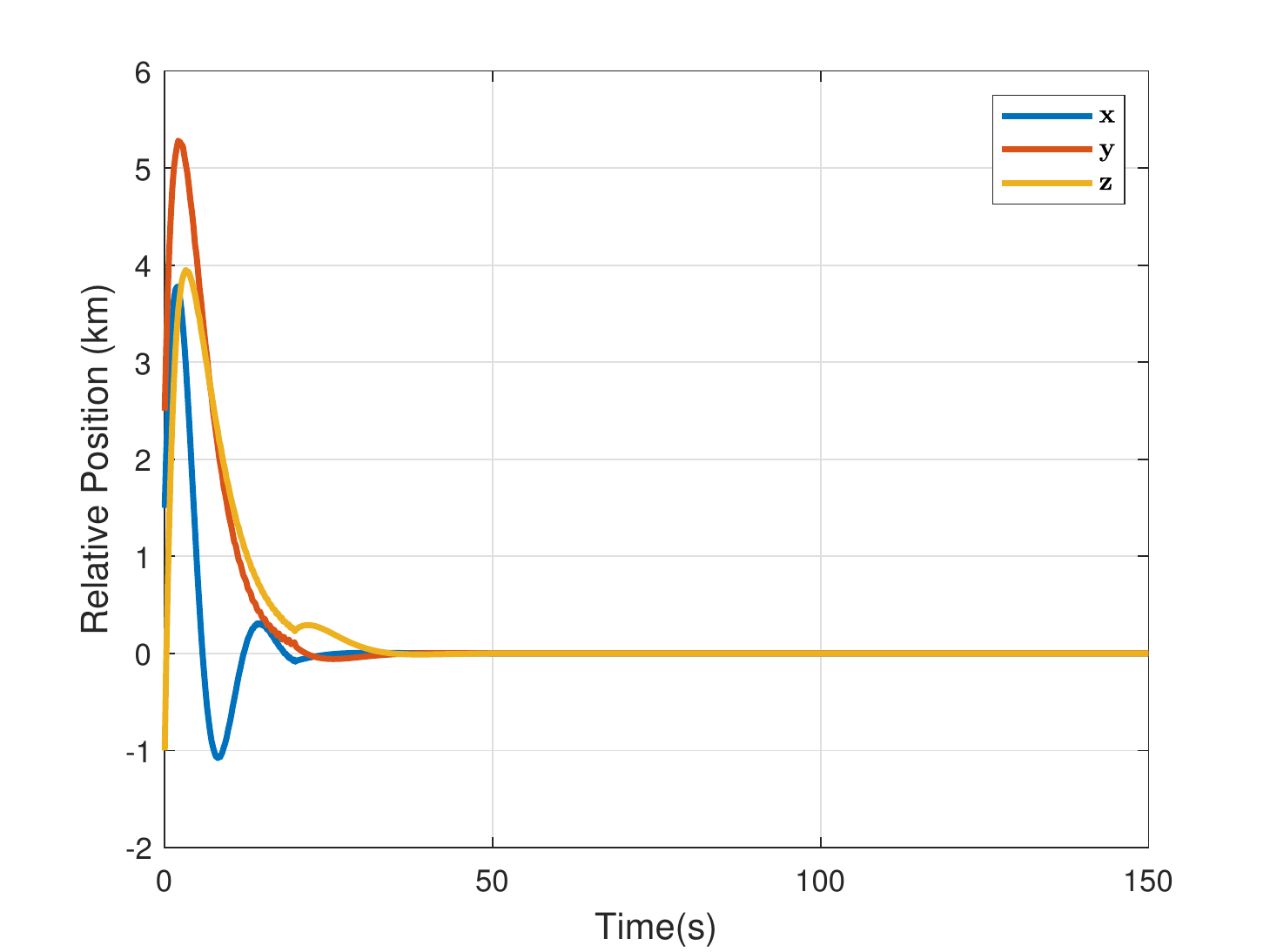}
    \caption{The relative positions $(\textbf{x},\textbf{y},\textbf{z})$ using the optimal control policy learned by Algorithm \ref{Algorithm:HI Data-driven}}
    \label{fig: relative position}
\end{figure}

\begin{figure}
    \centering
    \includegraphics[width=0.8\linewidth]{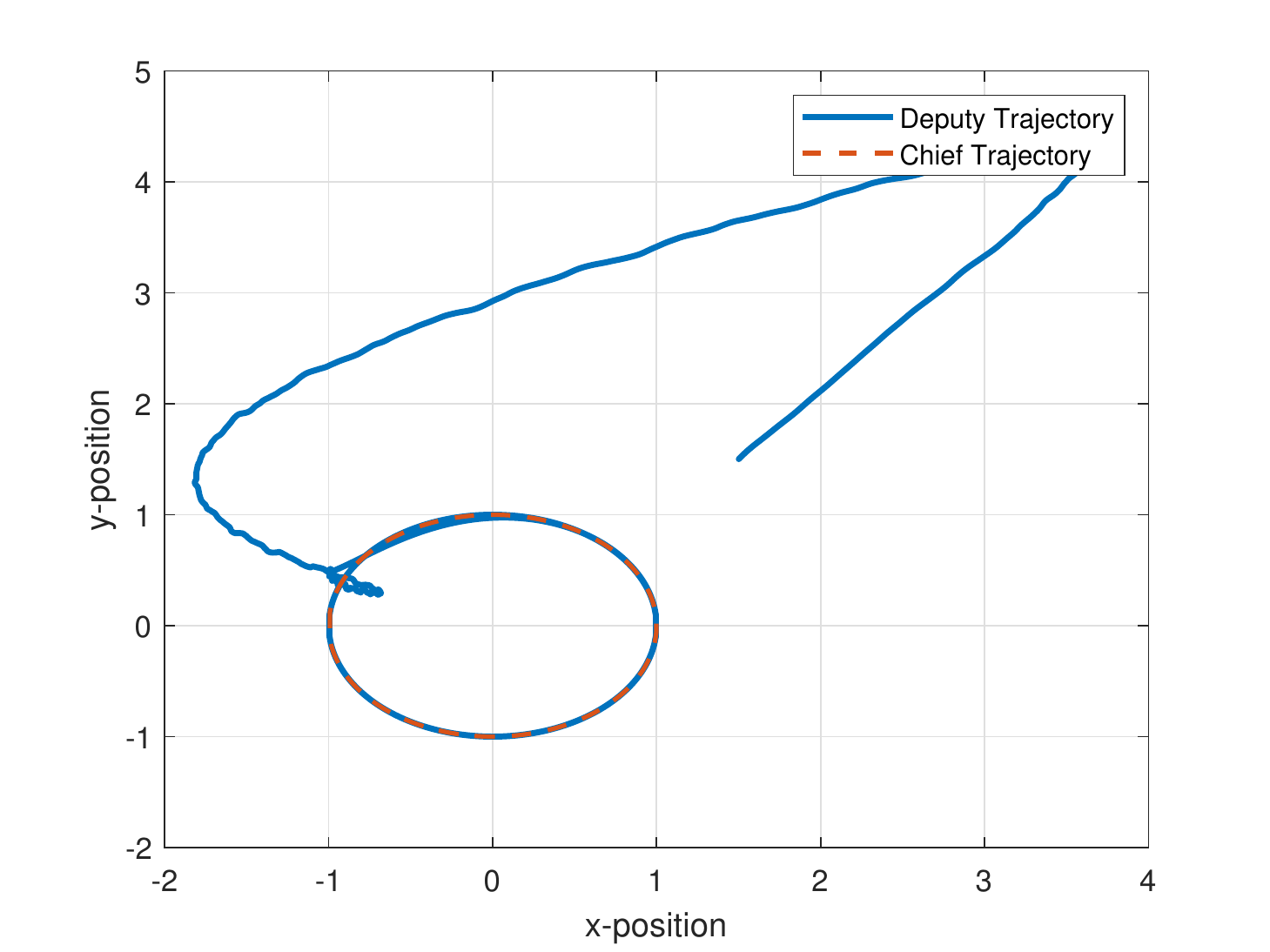}
    \caption{y-position vs. x-position while fixing the z-position using the optimal control policy learned by Algorithm \ref{Algorithm:HI Data-driven}}
    \label{fig: position xy}
\end{figure}

\begin{figure}[H]
    \centering
    \includegraphics[width=0.8\linewidth]{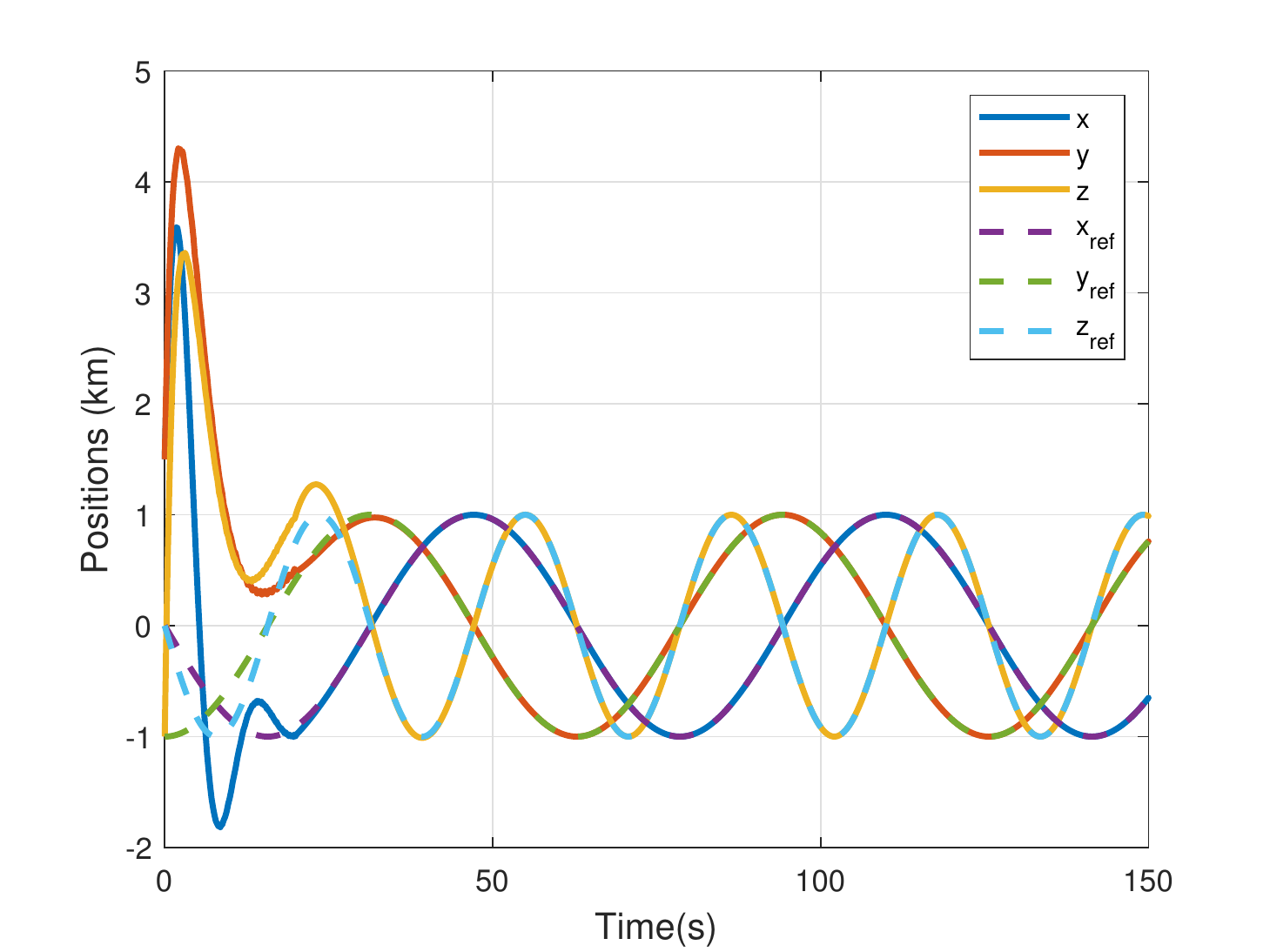}
    \caption{Deputy's position (x,y,z) and Chief's position $\text{(x}_\text{ref},\text{y}_\text{ref},\text{z}_\text{ref})$}
    \label{fig: position}
\end{figure}

For simulation purposes, we assume that the chief is moving in the space in all $x$, $y$, and $z$ directions. In addition, the velocity in each direction is different. Using Algorithm \ref{Algorithm:HI Data-driven}, the results are obtained depicted in Figs. \ref{fig: errorP}-\ref{fig: position}. In the following, the actual and the learned feedback and feedforward control gain matrices are shown. 

\begin{align*}
   L^{(35)}&=\begin{bmatrix}
  -2.1645 &  -4.0389 &   0&   0&       0.0005 &        0 &        0  &       0\\
    4.0387 &  -2.1644 &   0.0061  &  0  & 0  &     0     &    0       &  0\\
   0 &   0 &   0.8375 &  -8.0807  &       0  &       0  &  0.003 &0
      \end{bmatrix} \\
   L^{\star}&=\begin{bmatrix}
  -2.1644 &  -4.0387 &   0&   0&       0 &        0 &        0  &       0\\
    4.0387 &  -2.1644 &   0  &  0  & 0  &     0     &    0       &  0\\
   0 &   0 &   0.8377 &  -8.0807  &       0  &       0  &  0 &0
      \end{bmatrix}  
\end{align*}

 It is noticed that the learned control gain matrices are close enough to the optimal actual ones. In addition, one can realize from Figure \ref{fig: position} that the position of the deputy follows the position of the chief and the error converges to zero which confirms the completion of the docking procedure. Figure \ref{fig: errorP} illustrates the convergence of the learned value matrix to the optimal one. It is noted that the convergence of the value matrix is done in 5512 iterations. The large number of iterations incurred by the VI
 is due to the sublinear convergence rate of VI \cite{Bian2016}. However, the VI has less computational complexity comparing to the PI \cite{jiang2012computational} which converges in a quadratic convergence rate, but it still needs a stabilizing control policy to converge.

\section{Conclusion}
This work considers the control of autonomous satellite docking by developing a direct adaptive optimal control using adaptive dynamic programming (ADP). More specifically, two problems are considered to guarantee that optimal tracking. First, the output regulation problem is solved to achieve asymptotic tracking and disturbance rejection. Second, we consider solving the output regulation problem by adaptive dynamic programming. Therefore, the states and dynamics information of the system are not needed in order to compute the optimal feedback-feedforward control policy. The ADP approach is implemented on an autonomous satellite docking problem by considering the Clohessy-Wiltshire equation with J2 perturbations, where the problem is reformulated into an adaptive optimal output regulation problem. The simulation results illustrate the efficacy of the proposed method.

% \section*{Appendix}

% An Appendix, if needed, should appear before the acknowledgments.

% \section*{Acknowledgments}
% This work was supported, in part, by the U.S. National Science Foundation under Grant CMMI-2138206.

\bibliographystyle{AAS_publication}   % Number the references.
\bibliography{references}   % Use references.bib to resolve the labels.

\end{document}